\documentclass[final,5p,times,twocolumn]{elsarticle}

\usepackage{graphicx}
\usepackage{dcolumn}
\usepackage{bm}
\usepackage{subfigure}  
\usepackage{verbatim}
\usepackage{natbib}
\usepackage{float}
\usepackage{color}
 \usepackage{ amsmath}
 \usepackage{amssymb}

\newcommand{\vmu}{\mbox{{\bm{$\mu$}}}}
\newcommand{\vphi}{\mbox{{\bm{$\phi$}}}}
\newcommand{\vc}{\mbox{{\bm{$c$}}}}

\journal{Acta Materialia}

\begin{document}

\begin{frontmatter}



\title{Deviations from cooperative growth mode during eutectoid transformation: insights from phase-field approach\\
{\normalsize\textcolor{red}{\textbf{Accepted in Acta Materialia on $\mathrm{6}^{\mathrm{th}}$ August, 2014}}}}


\author[HSKA,KIT]{Kumar Ankit\corref{cor1}}
\ead{ankit.kumar2@hs-karlsruhe.de}
\author[HSKA,KIT]{Rajdip Mukherjee}
\author[KIT]{Tobias Mittnacht}
\author[HSKA,KIT]{Britta Nestler}
\cortext[cor1]{Corresponding author. Tel.:  +49 721 608-45022.}
\address[HSKA]{Institute of Materials and Processes, Karlsruhe University of Applied Sciences, Moltkestr. 30, 76133 Karlsruhe, Germany}
\address[KIT]{Institute of Applied Materials - Reliability of Components and Systems, Karlsruhe Institute of Technology, Haid-und-Neu-Str. 7, 76131 Karlsruhe, Germany}

\begin{abstract}
The non-cooperative eutectoid 
transformation relies on the
presence of pre-existing cementite particles
in the parent austenitic phase
and yields a product, popularly 
known as the divorced
eutectoid. In isothermal conditions,
two of the important parameters,
which influence the transformation mechanism
and determine the final morphology are
undercooling (below A$_{1}$ temperature) and
inter-particle spacing. Although, 
the criteria which governs
the morphological transition from lamellar to divorced
is experimentally well established, numerical 
studies that give a detailed exposition
of the non-cooperative transformation mechanism,
have not been reported extensively. 
In the present work, we employ a 
multiphase-field model, 
that uses the thermodynamic 
information from the
CALPHAD database,
to numerically simulate the 
pulling-away 
of the advancing ferrite-austenite interface
from cementite, which results in a transition from 
lamellar to divorced eutectoid morphology
in Fe-C alloy. 
We 
also 
identify 
the onset of a concurrent 
growth and coarsening 
regime at small 
inter-particle spacing and 
low undercooling.
We analyze the simulation results to 
unravel the essential physics behind this
complex spacial and temporal 
evolution pathway and 
amend the existing
criteria by constructing a 
Lamellar-Divorced-Coarsening (LDC) map.
\end{abstract}


\end{frontmatter}


\section{Introduction}\label{sec:Introduction}
\begin{figure}[htbp]
\centering
\includegraphics[width=0.45\textwidth]{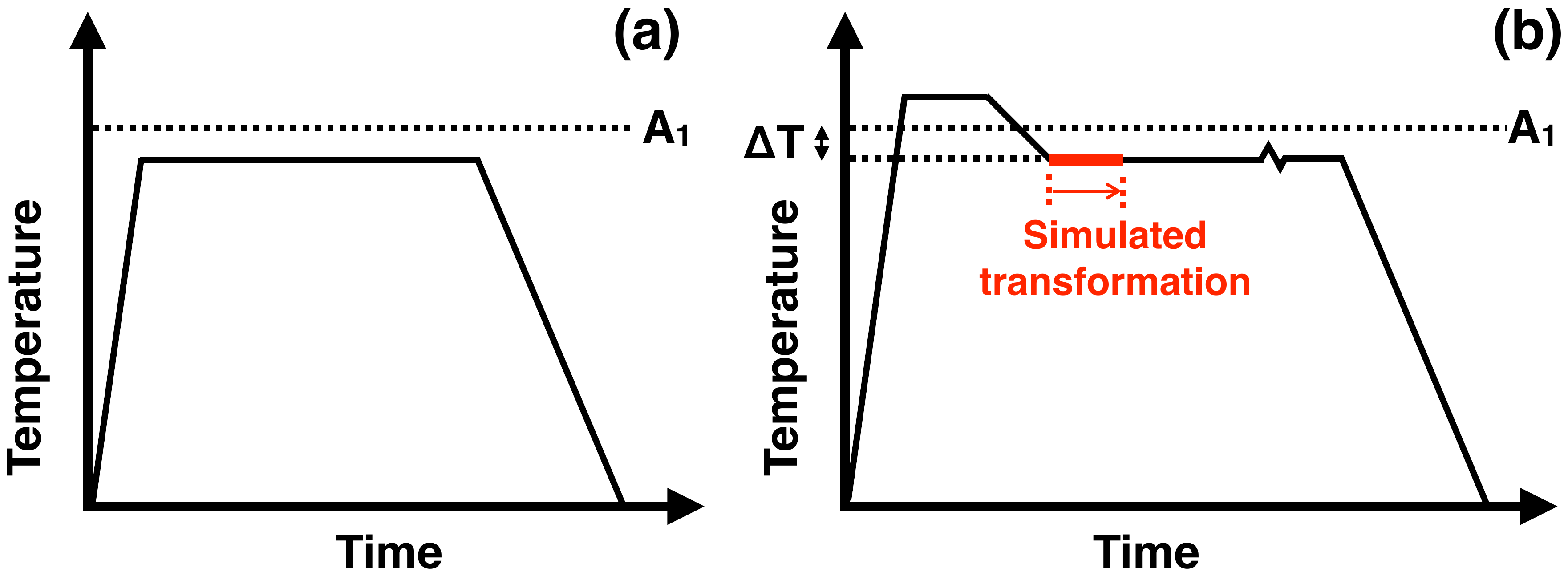}
\caption{Typical spherodizing 
annealing heat treatment 
cycles \cite{OBrien:1997kq}.
(a) Sub-critical annealing is
carried out slightly below the
A$_{1}$ temperature and 
does not involve the
formation of austenite.
(b) Inter-critical annealing involves heating the 
hypereutectoid steel to fully austenise it, with 
a small amount of cementite remaining undissolved and then, 
holding it just below
A$_{1}$ temperature. The final
transformation product is known as the
divorced eutectoid. The divorced eutectoid
transformation, that is numerically simulated
in the present work (for three different
undercoolings, $\Delta T$), is shown by the
colored (thick) line.
}
\label{fig:heat_treatment}
\end{figure}
\begin{figure}[tb]
\centering
\subfigure{\includegraphics[width=0.235\textwidth]{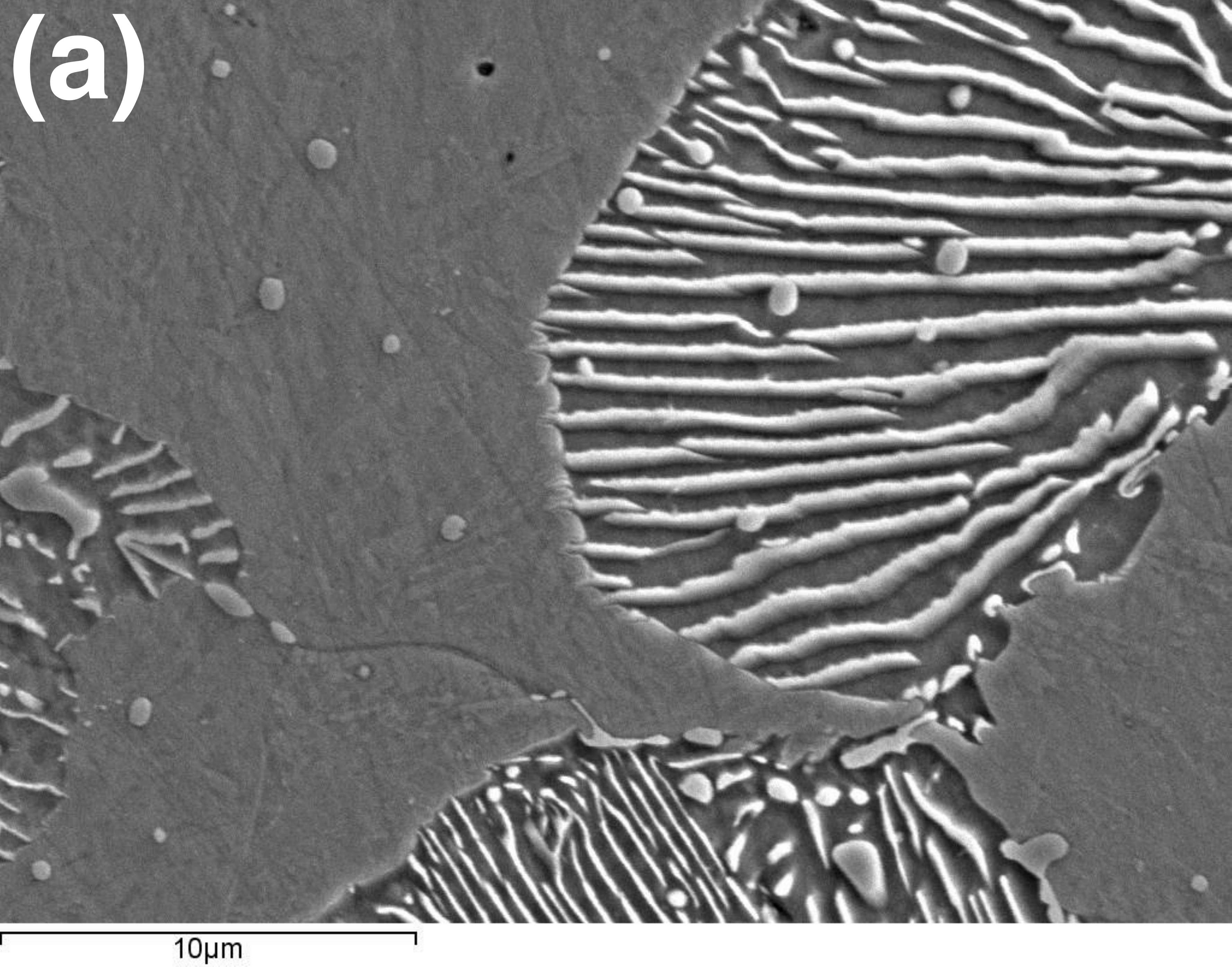}\label{fig:exp_lamellar}}
\subfigure{\includegraphics[width=0.235\textwidth]{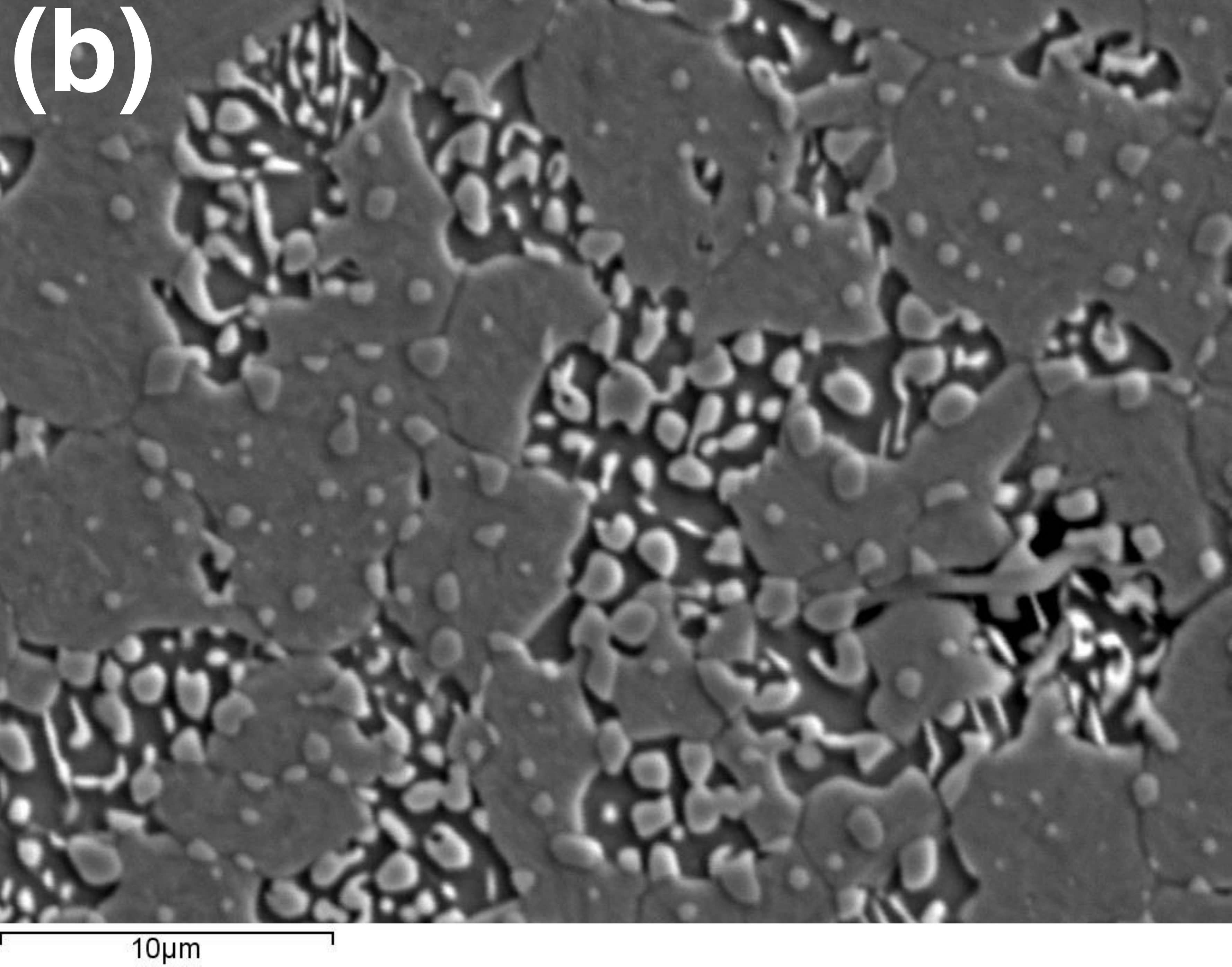}\label{fig:exp_divorced}}
\caption{Cooperative and non-cooperative growth 
regimes are observed during eutectoid 
transformation in 
Fe-0.92C-0.66Si-1.58Mn-1.58Cr-0.12Ni-0.05Mo-0.178Cu (wt. $\% $) alloy. 
Samples are austenized at $870\,^{\circ}\mathrm{C}$ for 
2.5 hours and held for $\sim 65$ minutes
below the eutectoid temperature
(at $\mathrm{T}_{1}$ and $\mathrm{T}_{2}$).
These are finally quenched to ambient temperature. 
(a) Lamellar and (b) divorced eutectoid 
morphologies are obtained for $\mathrm{T}_{1} = 710\,^{\circ}\mathrm{C}$
and $\mathrm{T}_{2} = 705\,^{\circ}\mathrm{C}$, respectively (private communication with Z.X. Yin and H.K.D.H. Bhadeshia).}
\label{fig:experimental_mic}
\end{figure}
The eutectoid transformation in steel involves
the decomposition of the parent austenite ($\gamma$) 
into two product phases, ferrite ($\alpha$-Fe)  and cementite  ($\theta$-Fe$_{3}$C).
When both the product phases, evolve cooperatively,
sharing a common growth front with austenite,
the morphology of the resulting product is lamellar, popularly known as
pearlite \cite{Zener:1947uq,Hillert:1957vn,Zackay:1962fk,Ankit:2013fk}. 
On the contrary, under a given set of conditions 
(low undercooling and small
inter-particle spacing of 
pre-exisiting cementite), the 
$\alpha/\gamma$ advancing 
transformation front begins to 
pull-away from cementite, 
leading to
the formation of a divorced eutectoid.
Hillert et al. \cite{Zackay:1962fk} establish that a
pearlitic colony comprises of inter-penetrating bi-crystals
of ferrite and cementite phases. Steels with a fully pearlitic 
microstructure (0.8 wt.$\%$ C),
find extensive application in the manufacture
of ropes, where high tensile strength is desirable.

Manufacture of a significant 
proportion of engineering components,
obligate the use of steels
with low hardness and good machinability (for e.g. in ball-bearings \cite{Bhadeshia:2012qy}).
Two, well-known 
spherodizing annealing 
heat treatment cycles,
that are adopted to
soften the pearlite, prior
to machining, 
are shown in 
Fig. \ref{fig:heat_treatment}.
The sub-critical annealing
involve the spherodization
of the fine pearlite,
by holding the 
hypoeutectoid steel
isothermally, just below 
the A$_{1}$ temperature,
as shown in Fig. \ref{fig:heat_treatment}(a).
The driving force
for morphological transition
is the reduction
in $\theta/\alpha$
interfacial area.
For softening hypereutectoid 
steels, intercritical annealing [Fig. \ref{fig:heat_treatment}(b)]
is a more economical method
(see \cite{Bhadeshia:2012qy}
and references therein). The steel is
fully austenised, such that a small 
amount of cementite
particles remain undissolved, 
and then held below A$_{1}$
to generate a spherodized 
transformation product 
(cementite particles
embedded in ferritic phase),
popularly known as \textit{divorced eutectoid microstructure}, 
which is much softer than the lamellar counterpart
i.e. pearlite [Fig. \ref{fig:exp_lamellar}].
Experimental studies indicate
that the 
presence of
pre-existing cementite
particles in the parent austenitic 
matrix results in the non-cooperation
between the ferrite and cementite phases 
\citep{Oyama:1984qy,Syn:1994kx,Taleff:1996yu}
and yields divorced eutectoid as the final
transformation product [Fig. \ref{fig:exp_divorced}].

The history of divorced eutectoid dates back to
the time of Honda and Saito \cite{Honda:1920yu}, who
report the morphological dependence of the
final-transformed product
(lamellar to completely spehrodized)
on the austenising temperature.
Oyama et al. \cite{Oyama:1984qy} describe 
a heat treatment schedule, that is adopted
for spherodizing a microstructure, comprising of
a mixture of pearlite and proeutectoid cementite.
Verhoeven and Gibson \cite{Verhoeven:1998vn}
develop a theoretical framework (for binary Fe-C alloy) 
to establish the criteria,
that governs the transition from lamellar to divorced 
eutectoid morphology.
Luzginova et. al \cite{Luzginova:2008kq} study
the influence of chromium concentration on the
formation of divorced pearlite in a hypereutectoid steel.
Pandit and Bhadeshia \cite{Pandit:2012fk}
amend the earlier theory of lamellar to divorced
eutectoid transition, by accounting for the diffusion of
carbon along the transformation front. 
 
It is apparent from the brief literature survey,  that much of
the investigation of divorced eutectoid transformation is primarily limited
to experimental and theoretical studies.
They delineate the basic concept of the evolution mechanism,
but unable to provide the finer details required
for tailoring the mictrostructure to achieve the desired
properties (e.g. better machinability).
Therefore, a theoretical understanding
of the complex evolution pathways
during the divorced eutectoid transformation
is paramount to comprehend 
the final microstructure, which 
is indispensable 
from a technological point of view.

In view of establishing a synergy between 
theoretical and experimental studies concerning the
eutectoid transformation, the
phase-field method holds great promise in terms of the
ability to describe the interface evolution in the
diffusion
length scale. In the present article, we use 
a multiphase-field model \cite{Choudhury:2012fe}
to scrutinize (and
amend) the existing theory by providing an 
in-depth understanding of the
carbon redistribution mechanism, 
which has profound implications in eventual
optimization of the process control parameters 
related to heat treatment of steel.
Based on the insights from numerical simulations, our further intention is to depict the interplay between 
two important parameters -- (a) spacing between the pre-existing cementite particles and 
(b) undercooling, which can result in different eutectoid morphologies. 

In the following section, the phase-field model,
used for the present numerical simulations, 
is briefly outlined. The 
simulation results concerning the lamellar
to divorced eutectoid transition and the concurrent growth and
coarsening regime are discussed in subsections \ref{sec:LD_transition} and \ref{sec:Coarsening} 
respectively. In subsection \ref{sec:LDC_map}, we summarize
the presented simulation results by constructing a 
lamellar-divorced-coarsening (LDC) map. 
Section 4 concludes the article.
\section{Phase-field model}\label{sec:Model}
The multiphase-field model is a
common diffuse-interface approach for 
studying microstructural evolution 
accompanying phase transformations.
The primary advantage of such a 
diffused-interface approach
lies in the elegance 
with which it treats 
moving boundary problems 
by obviating the necessity 
to explicitly track the position of interfaces.
In the present work, we use this approach
for numerical simulations, which is coupled 
with CALPHAD database to study 
a binary Fe-C alloy system. 
The multiphase-field
model equations,
that are used in the present study, 
is briefly outlined in this
section. 
The reader is referred 
to the previous  
studies \cite{Choudhury:2012fe,Molnar:2012fk,Ankit:2013fk,Mukherjee:2013kx}  
for a more detailed 
description of the 
model equations 
and numerical methods.

The evolution 
of phases is governed by
the phenomenological
minimization of the grand
potential functional $\Omega$,
\begin{align}
 {\Omega}&\left(T,\vmu,\vphi\right)=\nonumber \\
 &\int_{V}\Bigg[\Psi\left(T,\vmu,
\vphi\right)+\left(\epsilon{a}\left(\vphi,\nabla\vphi\right) +
\dfrac{1}{\epsilon}{w}\left(\vphi\right)\right)\Bigg]dV,
 \label{GrandPotentialfunctional}
 \end{align}
 where $T$ is the temperature, $\vmu$ is the 
  chemical potential vector comprising 
 of $K-1$ independent chemical potentials, 
 $\vphi$ is the phase-field vector
  containing the volume fractions of the 
  N-phases and
 $\epsilon$ is the length scale 
 related to the interface. ${a}\left(\vphi,\nabla\vphi\right)$ and
 ${w}\left(\vphi\right)$ represent the gradient 
 and obstacle potential type energy density, respectively
 and $V$ represents the domain volume.
The grand potential density 
$\Psi\left(T,\vmu,\vphi\right)$, 
which is the 
\emph{Legendre transform} of the
free energy density of the system 
$f\left(T,\vc,\vphi\right)$ is written
as an interpolation of
individual grand potential densities
\begin{align}
&\Psi\left(T,\vmu,\vphi\right) =
\sum_{\alpha=1}^N\Psi_{\alpha}\left(T,\vmu\right)h_{\alpha}\left(\vphi\right) \nonumber\\ 
&\Psi_{\alpha}\left(T,\vmu\right) =
f_\alpha\left(\vc^\alpha\left(T,\vmu\right),T\right) - \sum_{i=1}^{K-1}\mu_{i}
c_{i}^{\alpha}\left(T,\vmu\right),
 \label{interpolation_grandchem}
 \end{align}
 where $h_{\alpha}\left(\vphi\right)$ is an interpolation function of the form
 $h_{\alpha}\left(\vphi\right)=\phi_{\alpha}^{2}\left(3-2\phi_{\alpha}\right)$.
 The evolution equation for the N phase-field variables can be written as,
 \begin{align}
 \tau&\epsilon\dfrac{\partial \phi_{\alpha}}{\partial t}=\epsilon \left(\nabla
 \cdot \dfrac{\partial {a}\left(\vphi,\nabla \vphi\right)}{\partial \nabla
 \phi_{\alpha}}- \dfrac{\partial{a}\left(\vphi,\nabla
 \vphi\right)}{\partial \phi_\alpha}\right) \nonumber\\
 &-\dfrac{1}{\epsilon}\dfrac{\partial
 {w}\left(\vphi\right)}{\partial \phi_\alpha}-\dfrac{\partial
 \Psi\left(T,\vmu, \vphi\right)}{\partial \phi_\alpha}- \Lambda,
 \label{equation_evolution_phi}
 \end{align} 
 where $\Lambda$ is the 
 Lagrange parameter to maintain 
 the constraint
 $\sum_{\alpha=1}^N \phi_\alpha =1$. 
The concentration fields are 
obtained by a mass conservation 
equation for each
of the $K-1$ independent 
concentration variables $c_i$. 
The evolution equation 
for the concentration fields can be derived as,
\begin{eqnarray}
 \dfrac{\partial c_i}{\partial t} &=& \nabla \cdot
\left(\sum_{j=1}^{K-1}M_{ij}\left(\vphi\right) \nabla \mu_j\right)
\end{eqnarray}
\begin{eqnarray}
 M_{ij}\left(\vphi\right) &=& \sum_{\alpha=1}^{N}M_{ij}^\alpha
g_{\alpha}\left(\vphi\right),
\label{Concentration_equation}
\end{eqnarray}
where each $M_{ij}^\alpha$ represents the mobility matrix
of the phase $\alpha$ (related to the diffusivity).
The function $g_{\alpha}\left(\vphi\right)$ is in general not 
the same as $h_{\alpha}\left(\vphi\right)$ which 
interpolates the grand potentials, 
however, in the present description, 
we utilize the same.
The thermodynamic data-fitting 
procedure to approximate 
the variation of the 
grand-potential 
of the respective 
phases as a function 
of chemical potential and
the relation of the numerical simulation parameters
with the corresponding quantities in the sharp-interface limit,
are explained in the previous work \cite{Ankit:2013fk}.  

\section{Results and Discussion}
\subsection{Lamellar to Divorced transition}\label{sec:LD_transition}
\begin{figure}[t]
\centering
\subfigure{\includegraphics[width=0.45\textwidth]{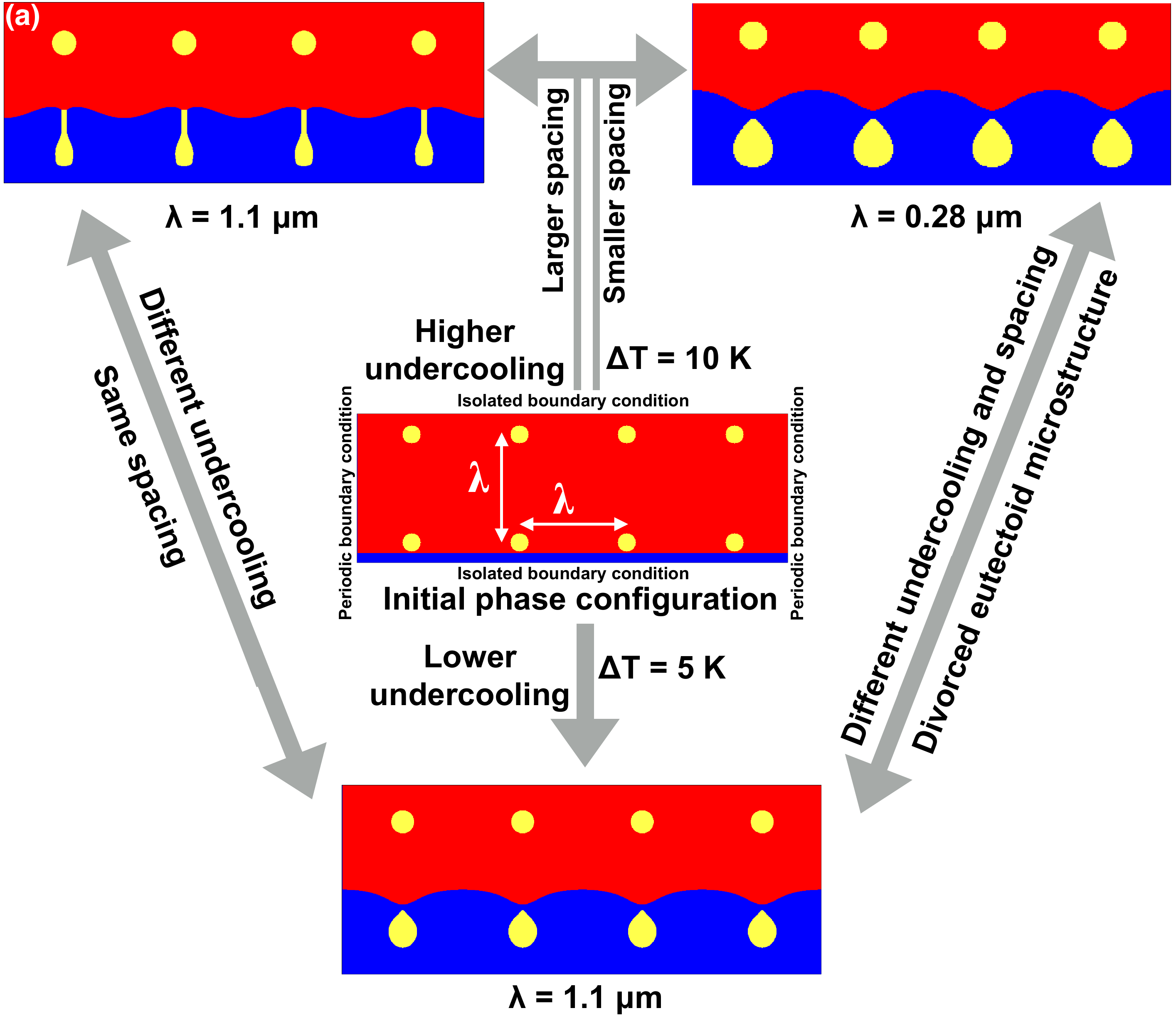}\label{fig:lam_div_transition}}\\
\subfigure{\includegraphics[width=0.4\textwidth]{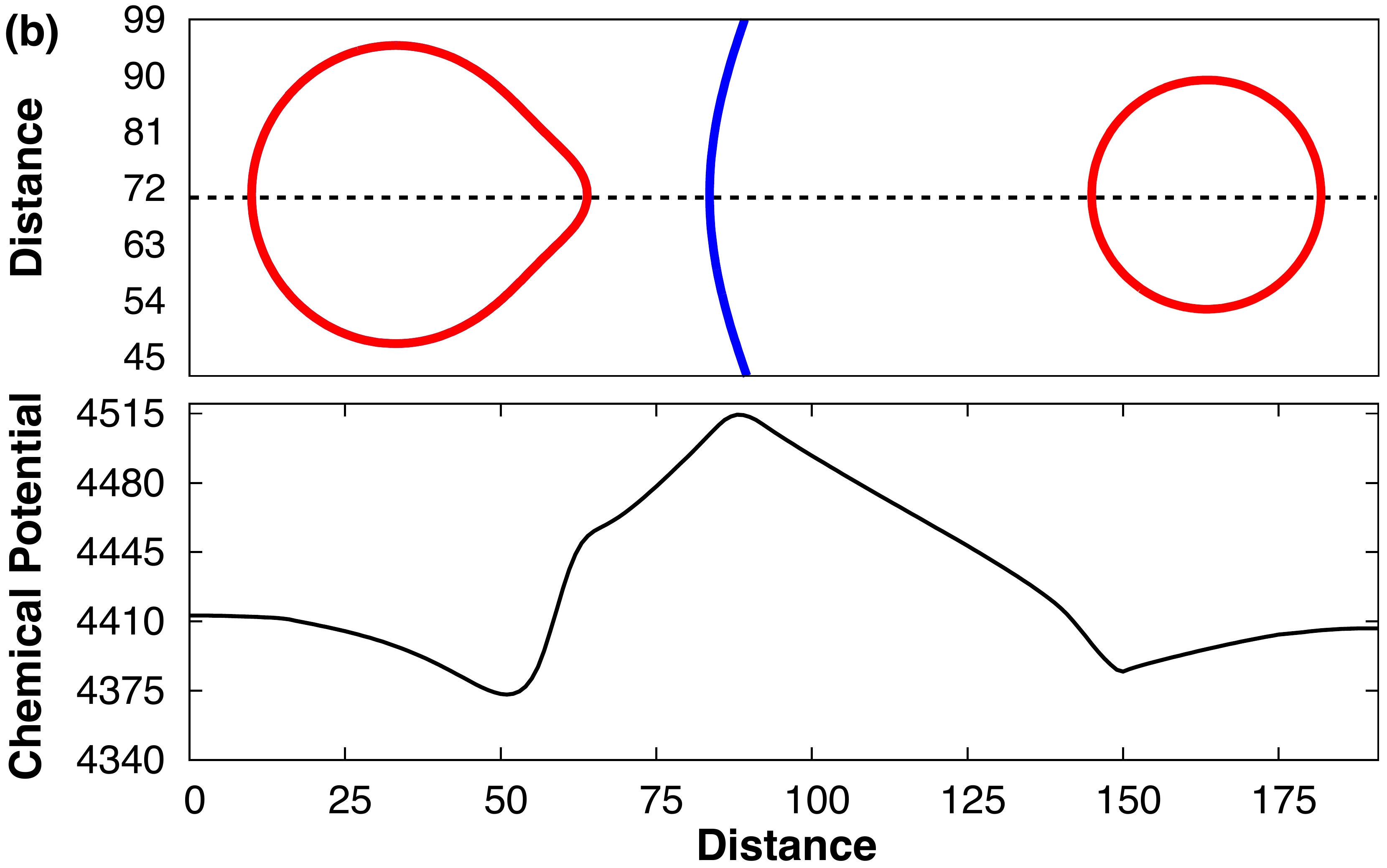}\label{fig:chempot_divorced}}\qquad
\caption{(a) Numerically simulated microstructures at two different
undercoolings below the eutectoid temperature
($\Delta T = 5\:\mathrm{and}\:10\:\mathrm{K}$) and particle spacing 
($\lambda = 0.28\;\mathrm{and}\:1.1\,\mu\mathrm{m}$)
starting from the same initial arrangement of the phases. The diagram 
shows that a cooperative growth regime is favored at
higher undercooling and spacing leading to the formation
of pearlitic lamellae. At lower undercoolings and smaller
  particle spacings, a non-cooperative mechanism predominates
  which results in the formation of a 
  divorced eutectoid microstructure.
  (b) 1-D chemical potential profile
  for $\Delta T = 5\mathrm{K}$ and $\lambda = 0.58\,\mu\mathrm{m}$
  plotted along the dashed line connecting the
  center of both the cementite particles.
  The profile shows that the
  carbon partitioned at the $\alpha/\gamma$
  transformation front is incorporated by
  both the particles which results in
  non-cooperative eutectoid transformation.}
\end{figure}
\begin{figure*}[!htbp]
\centering
\includegraphics[width=0.7\textwidth]{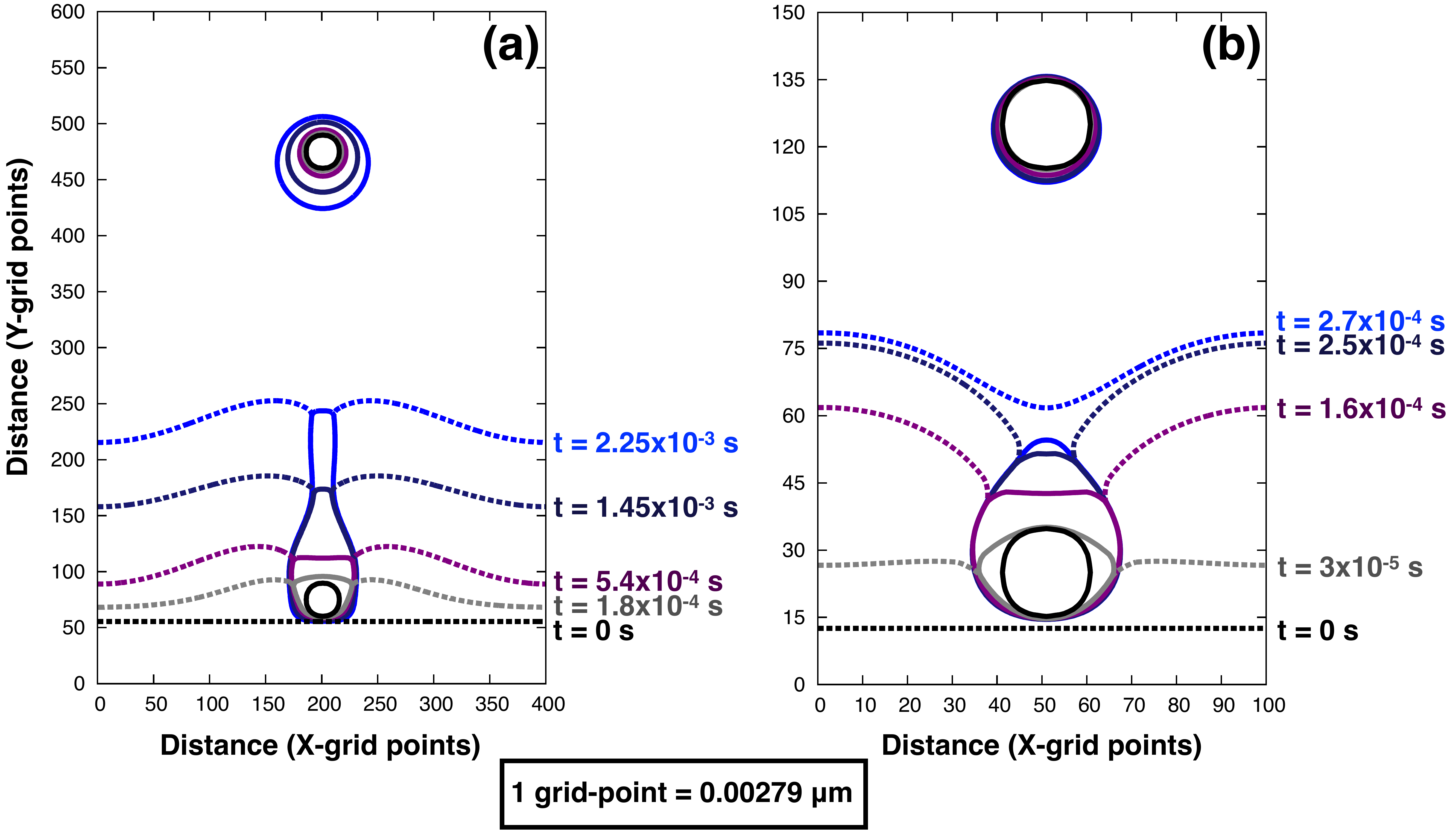}
\caption{Temporal evolution of the isolevels $\phi_{\alpha}=0.5$ 
(dashed lines representing $\alpha/\gamma$ interface)
and $\phi_{\theta}= 0.5$ (solid lines representing $\theta/\gamma$
and $\theta/\alpha$ interfaces) for the
(a) cooperative (resulting in pearlitic lamella) and 
(b) non-cooperative (resulting in divorced eutectoid) regimes. 
The pulling-away of the advancing ferrite-austenite
interface is evident from the numerically simulated 
isolevels shown in (b). A comparison of the 
temporal evolution of the isolines 
in (a) and (b) indicate that the initial particle spacing `$\lambda$'
(number of grid-points along x-axis) governs
the switch between both the 
evolution regimes (for constant undercooling 
$\Delta\mathrm{T} = 10\mathrm{K}$).}\label{fig:contour_lam_div}
\end{figure*}
As we are primarily interested 
in amending the 
criteria which determines
whether the
eutectoid transformation
front 
evolves by a
cooperative
(lamellar growth,
which leads to the
formation of
pearlite) or a
non-cooperative mechanism
(resulting in divorced eutectoid),
we use the same input parameters
(volume diffusion constants
and surface energies)
for the present phase-field simulations 
that was used earlier by Ankit et al. 
\cite{Ankit:2013fk} to simulate 
a pearlitic morphology.
In order to account for
the role of
diffusion of carbon 
along the transformation front
simultaneously, the
interface diffusion 
constant is
assumed to be 1000 times greater than 
volume diffusion constant in ferrite.
The interface
relaxation coefficient 
is derived from a 
thin-interface analysis
which is 
described elsewhere
\cite{Karma:1996kq,Choudhury:2012fe}.

We study the temporal
evolution of
austenite, ferrite and cementite phases
which is governed by the initial
particle spacing at intercritical 
temperature and the undercooling below
the eutectoid temperature ($A_{1}$).
The simulation domain width 
in the transverse direction
directly controls the spacing
(represented by $\lambda$) while the
radius of the particles
is kept same
for consistency of the
numerical results. 
In order to compare
the present phase-field
results 
with the classical theories,
which introduce a criteria 
for lamellar to divorced transition
based on experimental findings
\cite{Verhoeven:1998vn,Bhadeshia:2012qy},
we limit the present
discussion to a
symmetric arrangement
of pre-existing cementite particles
which  are embedded in an austenite
matrix. The undercooling below 
the eutectoid temperature ($\Delta T$) 
as well the particle spacing ($\lambda$)
is varied
to study their effect on the resulting microstructure.

Fig. \ref{fig:lam_div_transition}
shows the dependence of
undercooling and particle
spacing in stimulating
a transition from lamellar
to divorced morphology.
It is noteworthy, that the numerical results
accentuate the experimental findings
which emphasize a greater tendency of the
ferrite-austenite interface to pull away from cementite
particles at low undercooling and small spacing. 
On the contrary, at larger spacing 
and higher undercooling,
a cooperative growth regime is favored
which results in the formation of pearlitic lamellae.
On analyzing the simulated chemical potential 
profile in 1-D as shown in fig. \ref{fig:chempot_divorced},
it is apparent that
a divorced morphology 
forms due to the incorporation of
partitioned carbon
(at the advancing $\alpha/\gamma$ transformation front)
into the existing cementite 
particles. Thus, a near overlap
of 
the present
simulation results with
the existing theory
demonstrate the general
capability of phase-field method 
in capturing the topological
changes during eutectoid
transformation.

Fig. \ref{fig:contour_lam_div} compares the
temporal evolution of the numerically
simulated isolines corresponding to 
interphase interfaces for a lamellar
growth [Fig.~\ref{fig:contour_lam_div}(a)] and divorced eutectoid
[Fig.~\ref{fig:contour_lam_div}(b)], starting from a symmetric
arrangement of cementite particles. As the undercooling
is constant for both the cases ($\Delta\mathrm{T}=10\mathrm{K}$),
the evolution mode (cooperative or non-cooperative) 
is determined by the initial particle spacing
`$\lambda$' (represented by 
X-axes in Fig.~\ref{fig:contour_lam_div}). 
At a lower value of `$\lambda$'
($0.27\,\mu$m),
the $\alpha/\gamma$ interface pulls-away from
$\theta$, more commonly known as, the non-cooperative growth.
However, at a larger value of `$\lambda$' ($1.11\,\mu$m), the growing phases, 
$\alpha$ and $\theta$ maintain a common transformation 
front, by evolving cooperatively.
\subsection{Concurrent growth and coarsening}\label{sec:Coarsening}
\begin{figure*}[htbp]
     \begin{center}
 \includegraphics[width=0.8\textwidth]{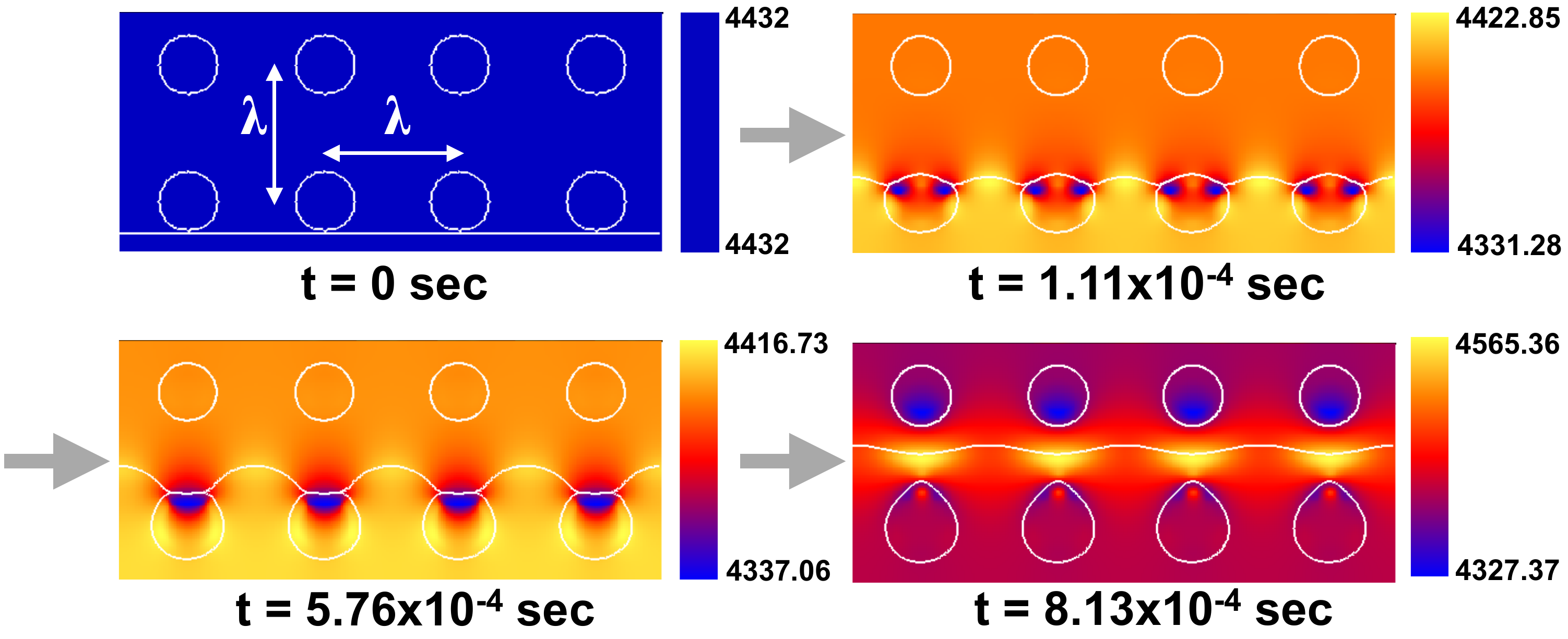}
     \caption{Temporal evolution of the phase contours 
     which are plotted over the corresponding
     chemical potential maps during concurrent
     growth and coarsening regime ($\Delta T = 7.5\:\mathrm{K}$
     and $\lambda = 0.294\:\mu{m}$). Coarsening
     can be observed clearly in Fig. \ref{fig:analysis_coarsening}.
     }
    \label{fig:chempot_coarsening}
     \end{center}
    \end{figure*}
\begin{figure}[!pt]
\begin{center}
\includegraphics[width=0.41\textwidth]{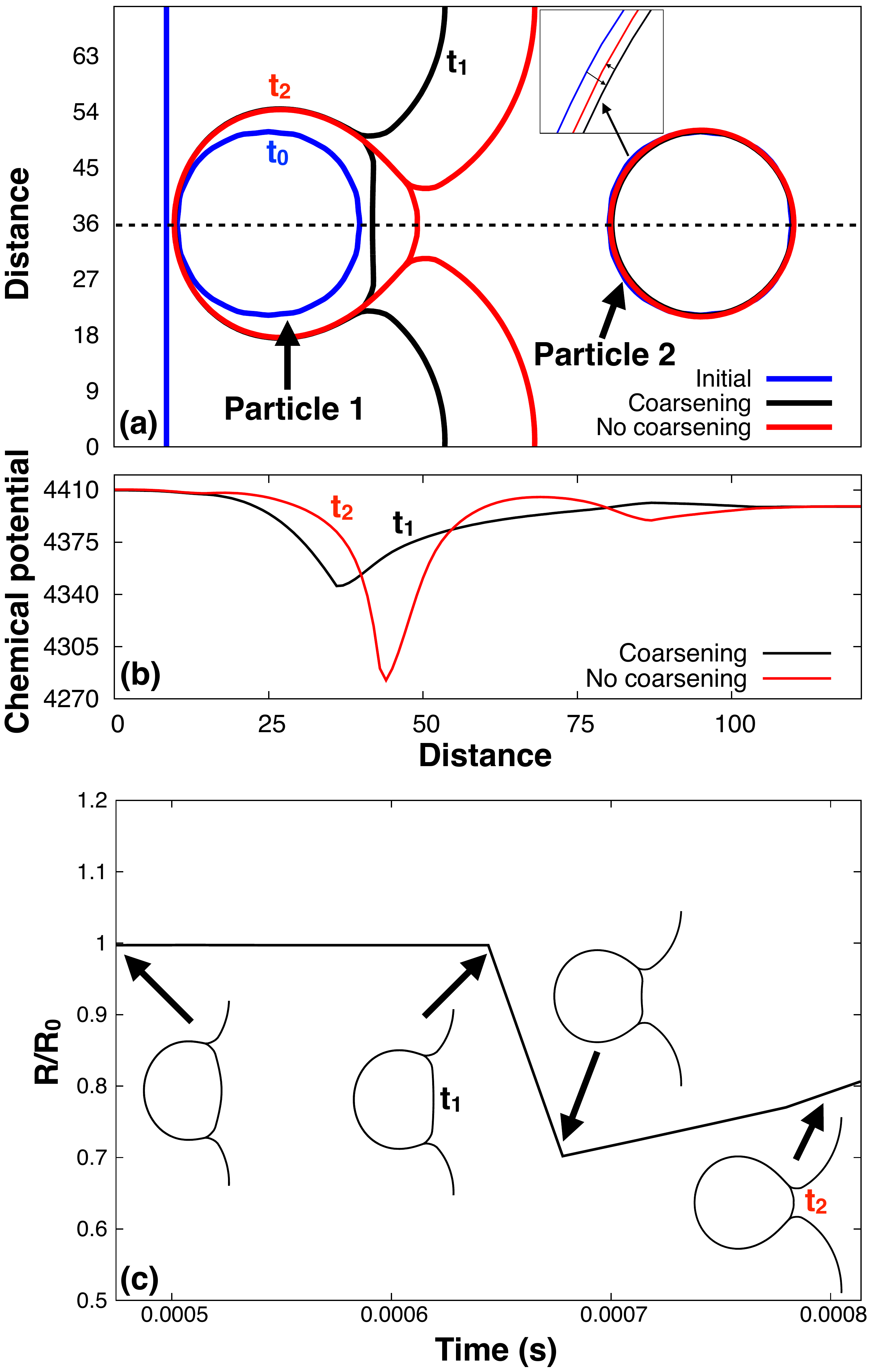}
\caption{
(a) Phase contours showing the subsequent 
shrinkage and growth 
of particle 2 as the $\alpha/\gamma$
front advances.
(b) 1-D chemical potential
profiles plotted along the dashed line in (a)
which shows a deviation in the carbon redistribution
mechanism during temporal evolution 
(as seen at $t_{1} = 6.44\times10^{-4}$
seconds and $t_{2} = 7.79\times10^{-4}$ seconds). (c) Temporal
evolution of the scaled radius ($R/R_{0}$) of particle 2
illustrating sharp deviations in the trend.
The corresponding contours of particle 1
are also plotted along side at different time-steps
which explains how the temporal change in curvature
of $\theta/\gamma$ interface of
particle 1 results in the onset of
growth and coarsening regimes respectively.}
\label{fig:analysis_coarsening}
\end{center}
\end{figure}
The most phenomenal finding
of the present numerical studies is the isolation
of concurrent growth and coarsening 
regime
during eutectoid transformation.
Fig. \ref{fig:chempot_coarsening}
shows the temporal
evolution of phase
contours which are
overlaid 
on the chemical potential
map, when the
initial cementite spacing
is reduced to 0.294 $\mu$m
at an undercooling of $5\:\mathrm{K}$ below the
$A_{1}$ (eutectoid) temperature.
An intermittent coarsening regime sets in 
before
the pulling away of the $\alpha/\gamma$ interface
from cementite particles.
In order to provide
a detailed exposition of this
newly identified regime
(which is not clearly visible in
Fig. \ref{fig:chempot_coarsening}), we
plot both the phase contours 
separately
[in Fig. \ref{fig:analysis_coarsening}(a)]
as well as the 1-D chemical
potential profile along the dashed-line
[in Fig. \ref{fig:analysis_coarsening}(b)]
for different 
simulation time-steps.
Depending on the initial 
distance from the
ferrite-austenite 
transformation front, 
the cementite particles are
labeled as 1
and 2.
On comparing the 1-D chemical
potential profiles for two different 
simulation time-steps ($t_{1}$ and $t_{2}$), 
we find that a change
in the carbon redistribution mechanism
is stimulated which
leads to coarsening of particles
prior to the divorce from the growth front. 

To begin with,
the $\alpha/\gamma$ 
transformation
front advances and
forms an interface with
the adjacent cementite particle 1.
As a result of this interaction, particle 1 starts
to grow due to the incorporation of 
partitioned carbon primarily via the interface (transformation front) 
diffusion flux. 
It is noteworthy,
that the particle 
1 which shares
a common interface with ferrite
experiences a greater influx of partitioned
carbon as compared to particle 2,
since the interface diffusivity is 
assumed to be 1000
times faster than the diffusion in austenite in all the
present cases. As 
the diffusion fields of both the
cementite particles
overlap, particle 1
grows while the particle 2
shrinks,
as shown in Figs. \ref{fig:analysis_coarsening}(a) and 
\ref{fig:analysis_coarsening}(c).
At this stage, the driving force for coarsening
predominates over the growth.
The same is also reflected [Fig. \ref{fig:analysis_coarsening}(a)]
by a temporal increase in the
curvature of $\theta/\gamma$ interface of particle 1
which slowly approaches infinity and subsequently
curves inwards.
An advancement of $\alpha/\gamma$
transformation front towards particle 2
causes a shift in the carbon redistribution
mechanism again; the driving force for
cementite growth exceeds coarsening.
We attribute a reduction in the distance
between $\alpha/\gamma$ transformation
front and particle 2 which makes the
incorporation
of partitioned carbon feasible
at smaller distances via bulk diffusion flux.
As a result, the chemical potential
near the advancing $\alpha/\gamma$
front ascends leading to the growth of particle 2.
This change in the carbon redistribution mechanism
which results in predominance of growth
over a coarsening regime
is evident from the 1-D plot 
shown in  Fig. \ref{fig:analysis_coarsening}(b).

It is worth clarifying that 
the ``concurrent
growth and coarsening" regime
(denoted by `C')
reported in the present work
principally differs
from the particle coarsening in alloys
which has been extensively reported
in the literature 
\cite{Mendoza:2004kk,Wu:2004qf,Ardell:2005fk,Clouet:2006xy,Hoyt:2011nr}.
Although, the reported regime `C' does 
involve curvature driven 
coarsening of particles, the primary difference
with the phenomena of conventional coarsening
is attributed to the energetics of $\alpha/\theta/\gamma$
phase triple-junction which 
determines if the 
transformation proceeds 
by a cooperative 
(to yield lamellar pearlite) 
or by a non-cooperative 
regime (yielding divorced eutectoid).
Further, the accompanying eutectoid 
transformation modifies the effective curvature
of $\theta/\gamma$ interface of particle `A'
which increases the rate of coarsening as depicted
by a decline in normalized radius of particle 'B' 
shown in Fig. \ref{fig:analysis_coarsening}(b).
It can be argued that the reported regime holds a
close resemblance with the 
discontinuous coarsening 
of grain boundary precipitates 
which could result in the formation 
of precipitate free zone (PFZ) 
along prior austenite grain boundaries \cite{Ahmadabadi:2011zr}.
However, on a careful examination, 
it is apparent that 
the physics of temporally 
evolving interphase interfaces
which is reported in the present study
is not only different, but also more
complex when compared to the
grain boundary interfaces 
involved in discontinuous coarsening.
\subsection{Lamellar-Divorced-Coarsening map (LDC)}\label{sec:LDC_map}
\begin{figure}[!htbp]
 \begin{center}
\includegraphics[width=0.3\textwidth]{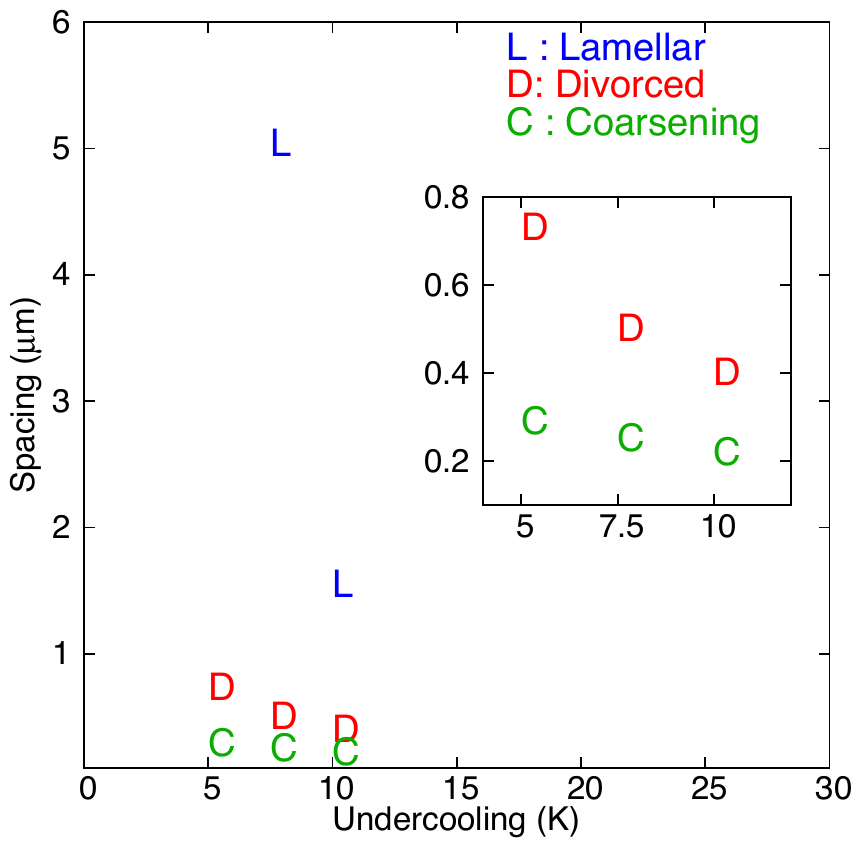}
 \caption{A morphological transition map showing the predominance of 
 lamellar (L), divorced (D) and concurrent growth and coarsening (C) modes
 during the eutectoid transformation
 in Fe-C alloy system. 
 The initial spacing between the cementite particles 
 (at the intercritical temperature) as well as the undercooling 
 below the $A_{1}$ temperature govern the 
switching among the three
 numerically simulated modes.
}
\label{fig:plot_LDC}
  \end{center}
 \end{figure}
 Having numerically
simulated and comprehended 
the ipseity of the concurrent
growth and coarsening regime
which precedes the non-cooperative
eutectoid transformation, we construct a
Lamellar-Divorced-Coarsening (LDC)
 transition map as shown in Fig. 
 \ref{fig:plot_LDC} to summarize
 the parametric study. The LDC
 transition map
 generated by conducting 
 phase-field simulations for three
 different undercoolings (5, 7.5 and 10 K)
below the eutectoid temperature 
and initial particle spacings
predicts the morphology that is
favored for a given set of
initial conditions during an isothermal 
transformation. 
In a nut-shell, the most 
significant contribution of the work presented
in the current letter is the addition of an alphabet
`C' (acronym for concurrent 
growth and coarsening regime
which is favored
at smaller spacing and lower 
undercooling)
to the classical Lamellar-Divorced (LD) map
\cite{Verhoeven:1998vn,Pandit:2012fk}. 
Further, the present numerical
findings are also in complete agreement 
with the existing theory for  
the divorced to lamellar morphological transition;
lamellar morphology being more favorable at large spacings
and high undercooling.
\section{Conclusions}\label{sec:Conclusion}
In conclusion, the 
spacing of the cementite particles
embedded in the austenite matrix
as well as undercooling below
the eutectoid temperature
entirely determines the final microtructure.
An in-depth phase-field study
of the isothermal transformation
presented in this article, aids
in selection of parameters to 
tailor the eutectoid microstructure
appropriately.
The present
approach also captures the
important transition between 
lamellar and divorced 
morphologies and sheds 
light
on the change in carbon redistribution
mechanism which is primarily 
governed by initial
configuration of the phases.
The concurrent growth and 
coarsening regime is
identified for
the first time which may be 
fundamentally difficult to isolate
in experiments. Thus, the present numerical
studies provide new insights into the transformation
mechanism 
and amend the classical model
of eutectoid transformation.

In future, it will be interesting to 
study the influence of asymmetrical
arrangement of cementite particles
on the final eutectoid morphologies.
Large-scale
numerical studies of the divorced 
eutectoid transformation 
for a random distribution of particles 
needs to be conducted to facilitate a
direct comparison with the
experimental microstructures.

\section*{Acknowledgements}
The authors thank Z.X. Yin and
Prof. H.K.D.H. Bhadeshia for the
contribution of experimental
microstructure and Prof. A. Choudhury
for preliminary discussions.
KA, RM and BN acknowledge the
financial support of DFG
in the framework of Graduate
School-1483.




%
%
%

%

%

%

%

%

%



\end{document}